\documentclass[iop,apj,tighten,twocolappendix,numberedappendix]{emulateapj}
\usepackage{apjfonts}

\usepackage{graphicx}
\usepackage{amsmath}
\usepackage{amssymb}
\usepackage{color}
\usepackage{mathtools}

\usepackage[breaklinks,colorlinks,citecolor=blue,linkcolor=red]{hyperref} 
\usepackage[all]{hypcap}

\newcommand\msun{\, \rm M_\odot}

\newcommand\be{\begin{equation}}
\newcommand\ee{\end{equation}}

\begin{document}

\title{An upper limit on the spin of SgrA$^*$ based on stellar orbits in its vicinity}

\author{Giacomo Fragione\altaffilmark{1,2}, Abraham Loeb \altaffilmark{3}}
 \affil{$^1$Center for Interdisciplinary Exploration \& Research in Astrophysics (CIERA), Evanston, IL 60202, USA} 
  \affil{$^2$Department of Physics \& Astronomy, Northwestern University, Evanston, IL 60202, USA}
  \affil{$^3$Astronomy Department, Harvard University, 60 Garden St., Cambridge, MA 02138, USA}

\begin{abstract}
The spin of the massive black hole (BH) at the center of the Milky Way, SgrA$^*$, has been poorly constrained so far. We place an upper limit on the spin of SgrA$^*$ based on the spatial distribution of the S-stars, which are arranged in two almost edge-on disks that are located at a position angle approximately $\pm 45^\circ$ with respect to the Galactic plane, on a milliparsec scale around the Galactic Center. Requiring that the frame-dragging precession has not had enough time to make the S-star orbital angular momentum precess, the spin of the massive BH at the center of the Milky Way can be constrained to $\chi\lesssim 0.1$.
\end{abstract}


\section{Introduction}
\label{sect:intro}

Central black holes (BHs) with masses $\sim 10^6\msun$--$10^9\msun$ are of fundamental importance in galaxy formation and evolution \citep{korm2013}. The impact of massive BHs in driving galactic outflows depends on their spin. The massive BH, SgrA$^*$, of approximately $4\times 10^6\msun$ at the center of the Milky Way, is the closest example. Its proximity enables high resolution observations of stellar orbits around it \citep{schodel2007}. This offers a unique opportunity for improving the understanding of galactic nuclei in general \citep*{genz2010}.

Within its sphere of influence, which encompasses nearly the central parsec of our Galaxy, SgrA$^*$ dominates the dynamics \citep{merritt2013,alexander2017}. The BH spacetime in general relativity is characterized by the BH mass and angular momentum. Despite the recent advancements in instrumentation with the Event Horizon Telescope \citep[EHT;][]{eht2019}, the intrinsic spin of the massive BH at the center of our Galaxy remains poorly constrained. The majority of previous work infers the spin by modeling interferometric EHT data sets with snapshot images of numerical simulations or semianalytic models \citep[e.g.,][]{dexter2010}. However, spin values deduced from the above approaches have been inconsistent, ranging from small values \citep{huang2009,brod2011,brod2016} to high values \citep{mosci2009,shch2012}. A robust constraint on the BH spin would have important implications, such as for its putative jetted emission \citep{falcke2004}.

\begin{figure} 
\centering
\includegraphics[scale=0.6]{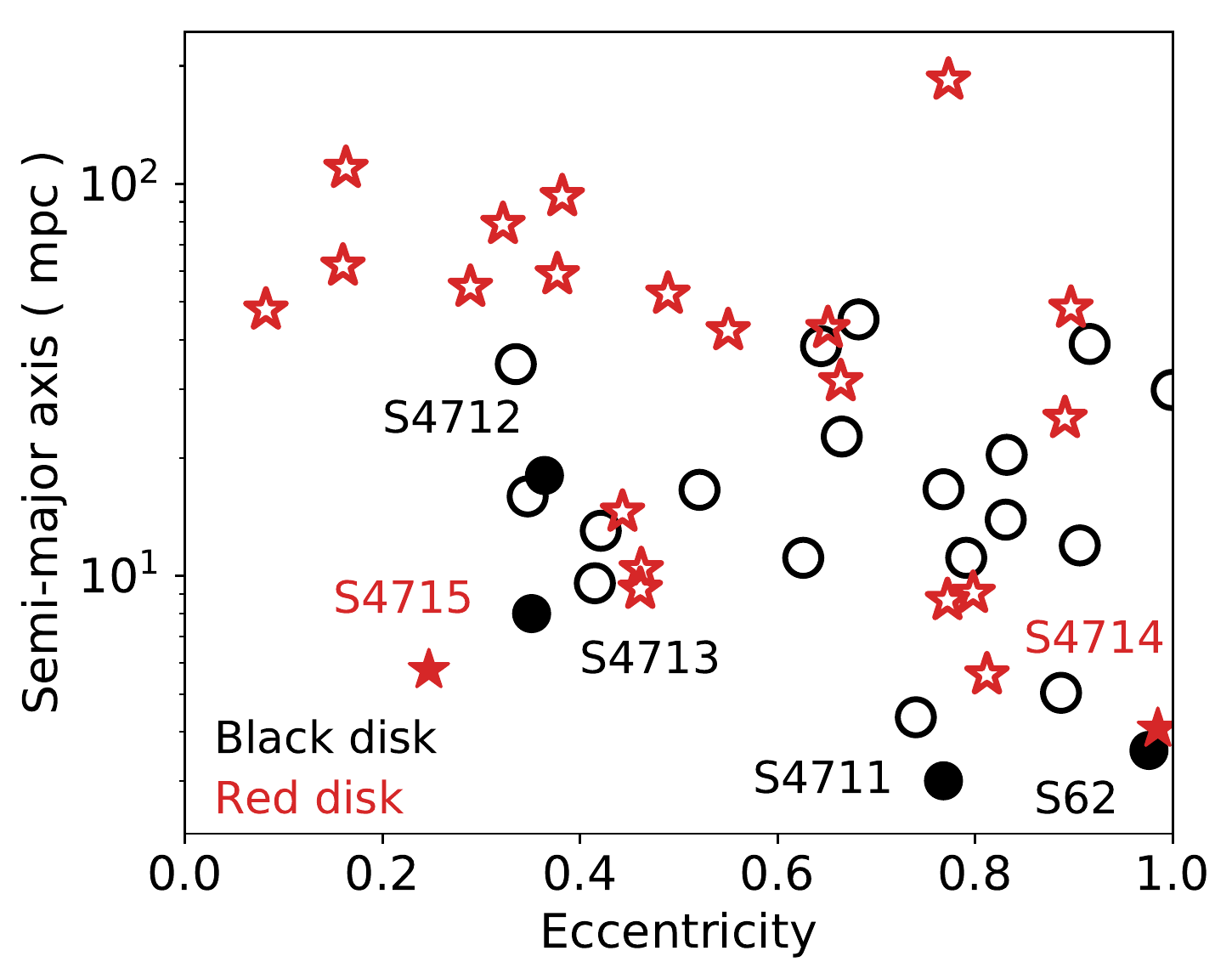}
\caption{Semi-major axis and eccentricity of the S-stars with known orbital parameters \citep{ali2020}. Red and black symbols show the S-stars in the ''red'' and ''black'' disks, respectively. Open symbols represent the classical population of S-stars \citep{gill2009,gill2017}, whereas filled symbols show the newly observed S-stars \citep{peiss2020a,peiss2020b}.}
\label{fig:observ}
\end{figure}

In this Letter, we argue that an independent constraint on the spin of SgrA$^*$ could be derived from the spatial distribution of the so-called S-stars, the closest stars to SgrA$^*$. In Section~\ref{sect:sstars}, we describe the orbital properties of S-stars. In Section~\ref{sect:spin}, we discuss how the motion of these stars can be used to put constraints on the BH spin. Finally, in Section~\ref{sect:concl}, we discuss the implications of our results and draw our conclusions.

\section{The fastest stars in the galaxy}
\label{sect:sstars}

\begin{figure*} 
\centering
\includegraphics[scale=0.65]{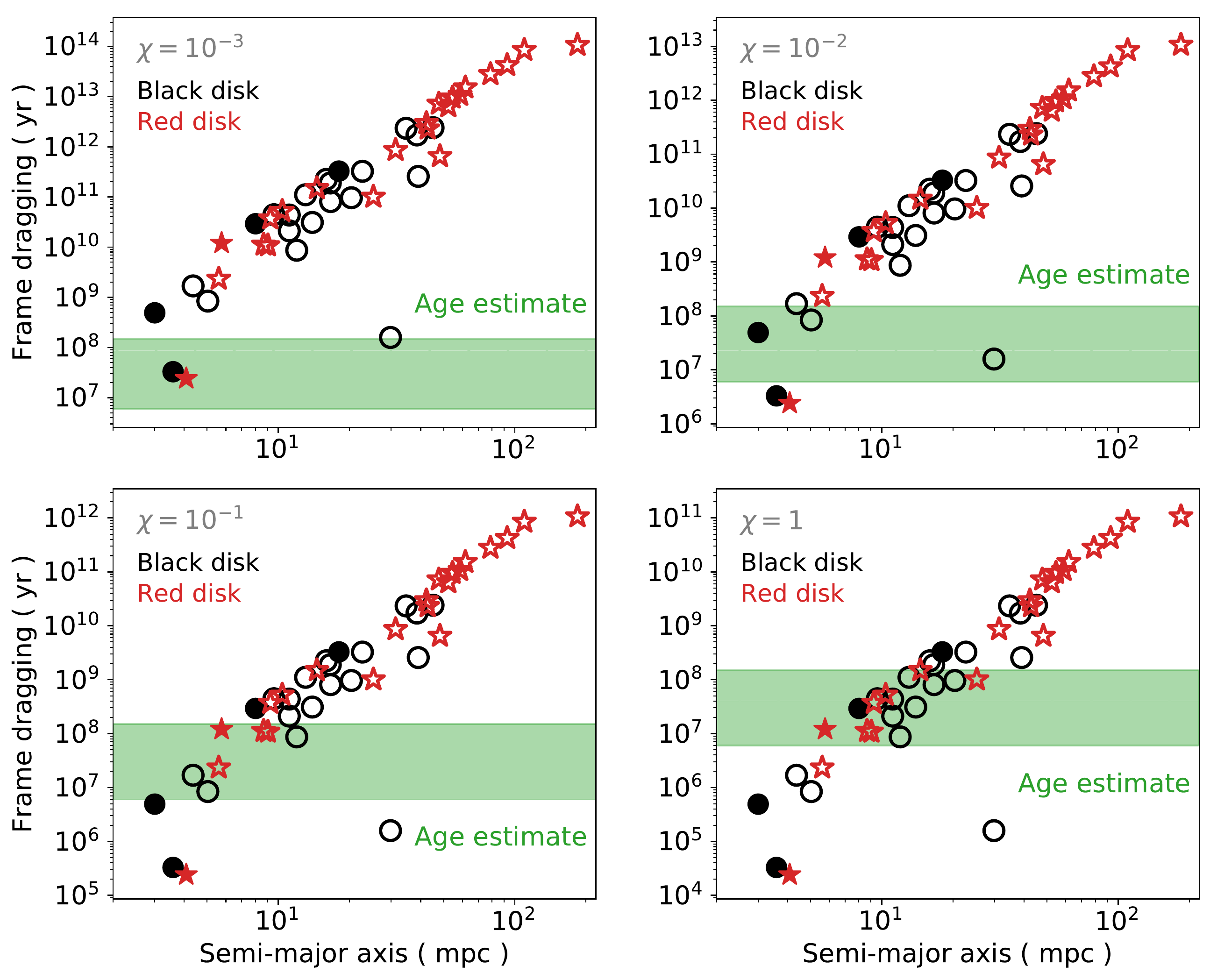}
\caption{Timescale for frame dragging (Eq.~\ref{eqn:ts}) as a function of the orbital semi-major axis of the S-stars with known orbital parameters \citep{ali2020}. Red and black symbols show the S-stars in the so-called red and black disks, respectively. Open symbols represent the classical population of S-stars \citep{gill2009,gill2017}, whereas filled symbols show the newly observed S-stars \citep{peiss2020a,peiss2020b}. The green region represent the estimated ages of S-stars \citep{habibi2017,peiss2020b}. Different panels correspond to different assumed values for the spin of SgrA$^*$. Top-left: $\chi=10^{-3}$; top-right: $\chi=10^{-2}$; bottom-left: $\chi=10^{-1}$; bottom-right: $\chi=1$.}
\label{fig:framdragg}
\end{figure*}

The S-stars, which constitute the stars in the innermost arcsecond of our Galaxy, are used as test particles under the influence of SgrA$^*$. Since the discovery of S2, with an orbital period of about $15$ yr \citep{sch2002Natur}, the number of stars with reasonably well-determined orbits has grown to about $40$ \citep{gill2009,gill2017}. The advent of the near-infrared GRAVITY instrument at the VLTI has marked the beginning of a new era in observations of the Galactic Center \citep{grav2018b,grav2018a}.

By combining data from NACO and SINFONI, \citet{ali2020} has recently shown that the S-stars are arranged in two nearly almost edge-on disks that are located at a position angle approximately $\pm 45^\circ$ with respect to the Galactic plane. In both disks, which they labeled ''black'' and ''red'' disks, the stars rotate in opposite directions.

Figure~\ref{fig:observ} shows the semi-major axis and eccentricity of the S-stars with know orbital parameters, as reported in Table~2 and Table~3 of \citet{ali2020}. We show the S-stars in the so-called red and black disks with red and black symbols, respectively. Void symbols represent the classical population of S-stars from \citep{gill2009,gill2017}.

Recently, six new S-stars have been discovered, some of them on orbits even closer to SgrA$^*$ than S2. The first of them has been S62, which travels on a $9.9$ yr orbit around the massive BH, was reported in \citet{peiss2020a}. The additional five stars, S4711-S4715, are fainter than the previously known S-stars, implying that they are less massive. While the classical S-stars have masses in the range $8\msun$--$14\msun$ \citep{habibi2017}, \citet{peiss2020b} estimated a mass of $6.1\msun$ for S62 and a mass in the range $2\msun$--$3\msun$ for S4711-S4715. They constitute a population of faint stars that can be found at distances to SgrA* comparable to the size of the solar system \citep{peiss2020b}. These six new stars are represented by filled symbols in Figure~\ref{fig:observ}. Four of them (filled black circles) are part of the black disk, while two of them (filled red stars) lie on the red disk.

Some of the S-stars have such small separations and pericenter passages from the central massive BH that they move at a fraction of the velocity of light. Therefore, S-stars constitute powerful tools for constraining the relativistic effects arising in the innermost regions of the BH spacetime.

\section{Upper limit on the spin of SgrA$^*$}
\label{sect:spin}

The spin angular momentum $\boldsymbol{\mathcal{S}}$ of a BH of mass $M_{\rm BH}$ can be expressed in terms of the dimensionless spin vector $|\boldsymbol{\chi}|$
\begin{equation}
\boldsymbol{\mathcal{S}}=\frac{GM_{\rm BH}^2}{c}\boldsymbol{\chi}\,,
\end{equation}
where $\chi=|\boldsymbol{\chi}|=0$ and $\chi=|\boldsymbol{\chi}|=1$ represent non-spinning and maximally-spinning BH, respectively. The BH spin induces a precession in the motion of a star, which is referred to as the Lense–Thirring or frame-dragging effect \citep{lt1918}. \footnote{The BH quadrupole moment induces a precession, usually negligible with respect to the precession induced by the BH spin.} The frame dragging precession affects the argument of periapsis, the longitude of the ascending nodes, and the orbital inclination \citep{iorio2020}. In particular, while the magnitude of the stellar orbital angular momentum does not change in time, its direction is torqued by the BH spin
\begin{equation}
\left\langle \frac{d\mathbf{L}}{dt}\right\rangle=\nu_{\mathcal{S}} \boldsymbol{\chi} \times \mathbf{L}\,
\end{equation}
where \citep[e.g.,][]{merritt2013}
\begin{equation}
T_{\mathcal{S}}=\frac{1}{\nu_{\mathcal{S}}}=\frac{c^2 a^3(1-e^2)^{3/2}}{2G^2M_{\rm BH}^2}
\label{eqn:ts}
\end{equation}
is the frame dragging precession timescale. In the previous equations, $a$ and $e$ are the orbital semi-major axis and eccentricity, respectively. Therefore, as a result of the frame dragging, the BH spin tends to align the stellar orbits to its equatorial plane. Note that the inclination of any stellar orbit within the BH equatorial plane would not be affected by frame dragging precession.

For our Galactic Center, frame dragging can have an appreciable effect on stellar orbits inside a milliparsec on timescales that are shorter than main-sequence lifetimes of massive stars \citep{levin2003}. As discussed in the previous section, some of the S-stars, in particular the newly found population \citep{peiss2020a,peiss2020b}, have very small pericenter passages from the central massive BH, where they move at a fraction of the speed of light. Therefore, the effect of the frame dragging precession could be significant if the BH spin is sufficiently high.

Figure~\ref{fig:framdragg} shows the timescale for frame dragging (Eq.~\ref{eqn:ts}) as a function of the orbital semi-major axis of the S-stars with known orbital parameters \citep{ali2020}. We use red and black symbols to label the S-stars in the red and black disk, respectively. As in Figure~\ref{fig:observ}, open symbols represent the classical population of S-stars \citep{gill2009,gill2017}, while filled symbols show the newly observed S-stars \citep{peiss2020a,peiss2020b}. The green region represent the estimated age for S-stars \citep{habibi2017,peiss2020b}. We plot the frame dragging timescale for different values of the BH spin: $\chi=10^{-3}$ (top-left), $\chi=10^{-2}$ (top-right), $\chi=10^{-1}$ (bottom-left), and $\chi=1$ (bottom-right). Since the chance to find aligned stars from a random distribution of orbital angular momentum is small, one could assume that most of the S-stars have formed in the same plane in which we find them today, through fragmentation of a thin gaseous disk. From Eq.~\ref{eqn:ts}, there would not be precession of the inclination of the stellar orbital planes lying on the BH equatorial plane. However, S-stars are arranged in two disks inclined about $90^\circ$ with respect to each other. As a consequence, in the extreme case one of them were to lay on the BH equator, implying no precession of its relative inclination, the frame-dragging precession would be maximal for the other disk of S-stars. In less extreme and more general spatial configurations, there would be a factor of order unity in the precession rate to account for the angle between the BH spin and the stellar angular momentum. Therefore, one cannot align the spin orientation to cancel out the blurring effect of both disks of stars. Since the frame-dragging timescale depends on the orbital semimajor axis and eccentricity of the S-stars, one cannot arrange for all the stars that were born in the same plane to change the orientation of their orbits at the same rate. As a consequence, the BH spin can be constrained demanding that the frame-dragging precession did not have enough time to make the S-star orbital angular momentum precess. This translates to requiring that $T_{\mathcal{S}}$ is larger than the typical age of S-stars. If this precession were to happen on a timescale shorter than the typical lifetime of S-stars, their configuration would not be a disk in space. Based on Figure~\ref{fig:framdragg}, this condition translates into $\chi\lesssim 0.1$. 

In summary, the orbital motion and spatial distribution of current S-stars limit the spin of the massive BH to $\chi\lesssim 0.1$.

\section{Conclusions}
\label{sect:concl}

Previous methods to constrain the spin of SgrA$^*$ provided inconsistent estimates that ranged from small values \citep{huang2009,brod2011,brod2016} to high values \citep{mosci2009,shch2012}. 

By requiring that the orbital planes in which the S-stars formed and are found today will not be erased by frame-dragging precession during their lifetime, we derived a strict upper limit on the spin of SgrA$^*$, $\chi\lesssim 0.1$.

\section*{Acknowledgements}

We thank Lorenzo Iorio for useful comments. GF acknowledges support from a CIERA Fellowship at Northwestern University. This work was supported in part by Harvard's Black Hole Initiative, which is funded by grants from JFT and GBMF.

\bibliographystyle{yahapj}
\bibliography{refs}

\end{document}